\documentclass[useAMS,usenatbib,usegraphicx]{mn2e}
\usepackage{amssymb}

\def\k{km s$^{-1}$}
\def\kms{km s$^{-1}$~}
\def\fant{$\phantom 0$}

\def\pp{^{\prime\prime}}

\def\cm2{cm$^{-2}$}
\def\pp{^{\prime\prime}}
\def\HII{H\,{\sc ii}}
\def\HI{H\,{\sc i}}
\def\mjb{mJy beam$^{-1}$}

\def\farcs{\hbox{$.\!\!^{\prime\prime}$}}

\def\ga{{\hbox{$\gtrsim$}}}
\def\deg{\hbox{$^\circ$} }
\def\fdeg{\hbox{$.\!\!{}^\circ$}}

\def\fs{\hbox{$.\!\!{}^{\rm s}$}}

\title[HI and radio continuum of G290.1--0.8]{High resolution HI and radio 
continuum observations of the SNR G290.1--0.8}
\author[Reynoso et al.]{Estela M. Reynoso$^{1,2}$\thanks{E-mail:
ereynoso@iafe.uba.ar}\thanks{Member of the Carrera del 
Investigador Cient\'\i fico, CONICET, Argentina.}, Simon Johnston$^{3}$, 
Anne J. Green$^{2}$ and B\"arbel S. Koribalski$^3$\\
$^{1}$Instituto de Astronom\'\i a y F\'\i sica del Espacio (IAFE),
C.C. 67, Suc. 28, 1428 Buenos Aires, Argentina\\
$^{2}$School of Physics, University of Sydney, NSW 2006, Australia\\
$^{3}$Australia Telescope National Facility, CSIRO, P.O. Box 76, 
      Epping, NSW 1710, Australia}

\begin{document}

\date{Accepted 2006 March 13. Received 2006 February 22}

\pagerange{\pageref{firstpage}--\pageref{lastpage}} \pubyear{2006}

\maketitle

\label{firstpage}

\begin{abstract}

We have observed the supernova remnant (SNR) G290.1--0.8 in the
21-cm \HI\ line and the 20-cm radio continuum using the Australia 
Telescope Compact Array (ATCA). The \HI\ data were combined with 
data from the Southern Galactic Plane Survey to recover the 
shortest spatial frequencies. In contrast, \HI\ absorption was
analyzed by filtering extended \HI\ emission, with spatial 
frequencies shorter than 1.1 k$\lambda$.
The low-resolution ATCA radio continuum image of the remnant 
shows considerable internal structure, resembling a network of 
filaments across its 13 arcmin diameter. A high-resolution ATCA 
radio continuum image was also constructed to study the small 
scale structure in the SNR. It shows that there are no structures 
smaller than $\sim 17\pp$, except perhaps for a bright knot to 
the south, which is probably an unrelated object. 
The \HI\ absorption study shows that the gas distribution and 
kinematics in front of SNR G290.1--0.8 are complex. We estimate 
that the SNR probably lies in the Carina arm, at a distance 
7 ($\pm1$) kpc. In addition, we have studied nearby sources in 
the observed field using archival multiwavelength data to 
determine their characteristics.

\end{abstract}


\section{Introduction}

The supernova remnant (SNR) G290.1--0.8 was originally discovered by 
\citet{msh61} in their survey of radio sources and named MSH 11-6{\it 1}A. 
It was first recognised as an SNR by \citet{kes67}. Based on observations 
at 408 MHz \citep{kc87} and at 5 and 8 GHz \citep{mck+89}, \citet{gre96} 
classified the remnant as shell-type with an angular size 
of $15^\prime \times 10^\prime$, a flux density $S$ of 40 Jy at 1 GHz, 
and a spectral index $\alpha$ ($S \propto \nu^\alpha$) of --0.4.
\citet{wg96} observed this SNR at 843 MHz with the Molonglo telescope
and found considerable complex internal structure rather than the 
well-defined annular brightness distribution typical of shell SNRs.

\citet{grbm72} observed the SNR in \HI\ with a spectral resolution of 2 
\kms using the Parkes radiotelescope. Their spectrum shows \HI\ emission 
in three broad peaks near --20, --10 and +25 \k, appearing to extend
to +50 \k. From the absorption spectrum they assigned a lower distance 
limit of 3.6 kpc (the tangent point using the earlier model of $R_\odot 
=$ 10 kpc) to the SNR. In a contemporary paper, \citet{dic73} claimed the 
SNR was {\it at} the tangent point but with little convincing evidence.

The optical counterpart to the radio SNR was discovered virtually
simultaneously by \citet{kw79} and \citet{em79}. Both papers
concluded that the remnant was much older and much further
away than assumed by \citet{dic73} 
and estimated the distance to be 12 to 15 kpc. More recently, 
\citet{ralm96} studied the kinematics of G290.1--0.8 based on H$\alpha$
observations and concluded that the kinematic velocity of the SNR is
+12 \k. However, the authors note the high complexity of the H$\alpha$
emission along the line of sight to the remnant.

\citet{sew90} detected the SNR in X-rays as part of his comprehensive
survey of Galactic SNRs. Although the X-ray morphology shows some
similarity to the radio shape, the emission is strongly peaked towards 
the centre. For this reason, Seward \nocite{sew90} classifies the remnant as 
filled-centre in contrast to the radio classification as a clumpy shell.
\citet{sshp02} analyzed {\it Advanced Satellite for Cosmology and 
Astrophysics} (ASCA) observations of G290.1--0.8 and found that 
the X-ray emission is of thermal origin. Therefore, this remnant should
be classified as a ``mixed-morphology'' SNR, which are shell-like
in radio wavelengths and centrally peaked in thermal X-ray emission
\citep[e.g.][]{rp98}. With such a model, \citet{sshp02} estimate
that the distance to G290.1--0.8 must be in the range 8-11 kpc, while 
the age should be between 10-20 kyr.

Shaver \& Goss (1970a,b)\nocite{sg70a,sg70b} published images 
of the remnant at 408 and 5000 MHz in a field containing two other strong 
sources, G289.9--0.8 and G290.4--0.9. The first of these sources has a 
thermal spectrum and they classified the second as extragalactic based 
on its spectral index and the fact that it was unresolved using a $3^\prime$
beam. \citet{dm72} classified G289.9--0.8 as an \HII\ region based on a 
detection of a recombination line with a velocity of +5 \kms and assigned a 
distance of 7.5 kpc to it. \citet{gcbt88} observed a CO cloud at $l=289\fdeg3,\ 
b=-0\fdeg6$ with a velocity of +22 \kms to which they assign a distance of 7.9 
kpc. Other \HII\ regions in this vicinity have similar velocities and
may be associated with the CO cloud. \citet{gcbt88} comment that the SNR is 
`seen close to a small CO peak'. IRAS and MSX images in this region show the 
\HII\ region G289.9--0.8 very clearly but little or no infra-red emission is 
seen from the SNR. As seen in the \citet{sg70a} and the \citet{gcbt88} images,
there is an extended and complex region around $l = 290\deg$ consisting
of many \HII\ regions and CO clouds which are clearly beyond the solar circle 
at a kinematic velocity of 22 \k . The SNR is located in the middle of 
this complex but there is no evidence of association.

Three pulsars lie in the vicinity of G290.1--0.8. Both PSR J1103--6025 
\citep{kbm+03} and PSR J1104--6103 \citep{kmj+96} have characteristic ages 
greater than one Myr and are located well outside the boundaries of the 
SNR. Of more interest is PSR~J1105--6107 \citep{kbm+97}, a rapidly 
rotating pulsar with a characteristic age of only 63 kyr. Its dispersion
measure is 271 cm$^{-3}$pc, which implies a distance of 7 kpc, although
the pulsar is located over 2 remnant radii away from SNR G290.1--0.8. 
Taking the age and distance at face value, and assuming the pulsar was 
born in the same event which produced the SNR, it would then need an average 
velocity of $\sim$700 \kms to reach the current position. This is high, 
although not impossible \citep[e.g.][]{chat05}. \citet{sshp02} and
\citet{jw06} suggest that PSR J1105-6107 is not likely to be associated 
with the remnant.

In this paper we present new high resolution \HI\ line and 20-cm radio 
continuum observations towards G290.1--0.8 to understand more about 
this SNR and its environment.

\section{Observations and data reduction}

Simultaneous \HI\ line and 20-cm radio continuum observations were carried 
out with the 1.5C and 750C configurations of the Australia Telescope Compact 
Array (ATCA) on 1995 May 16 and June 1, respectively. Both observations were 
of 14 hr duration.
With these configurations combined, there are 20 baselines ranging from 46 to 
1480 m and an additional 10 baselines between 2000 and 5000 m. 
The phase centre was at (J2000) RA $=11^{\rm h}\,03^{\rm m}\,24\fs0$, Decl.= 
--60\degr\,56\arcmin\,11\farcs0. The primary beam has a FWHM of 33\arcmin\
at 1400 MHz. The \HI\ measurements were centred at 1421 MHz using a 4 MHz 
bandwidth which was divided into 1024 channels. The velocity resolution is 1 
\k . The radio continuum measurements were centred at 1384 MHz using a 
bandwidth of 128 MHz. The data reduction 
was carried out with the {\sc miriad} software package \citep{stw} using 
standard procedures. The primary flux and bandpass calibrator was PKS
1934--638, while PKS 1036--697 and PKS 1215--457 were used for phase 
calibration. 

\subsection{Radio continuum data}

A low-resolution radio continuum image was constructed using all 
visibilities corresponding to baselines shorter than 1480 m. The data 
were Fourier-transformed with `uniform' weighting to minimize sidelobe 
levels. Compact sources were cleaned first, and then the residual image 
was used to find the clean components corresponding to diffuse emission. 
All clean components were finally convolved with a beam of $25\pp 
\times 23\pp$, with a position angle (P.A.) of $-47\deg$. The 
final image, shown in Figure~1, has an rms noise level of $\sim 1$ \mjb .

We also made a high resolution image of this region using only the 
visibilities from 12.5 to 24 k$\lambda$ (baselines from 2.6 to 5 km). In
this {\it u-v} data set, structures larger than $\sim 17\pp$ are filtered 
out, and we find that virtually all the emission from the SNR is resolved 
out. All continuum images were corrected for the primary beam response.

Using the capability of ATCA to measure simultaneously continuum emission 
from two orthogonal linear polarizations, we constructed images of the 
Stokes parameters Q and U. No polarized emission was detected, and we do
not discuss this further.

\begin{figure} 
\includegraphics[width=252pt]{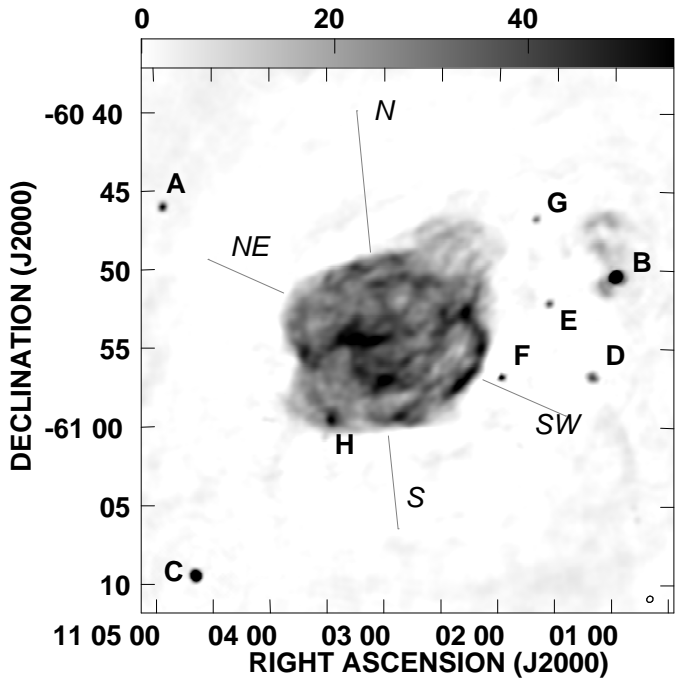}
\caption{Radio continuum image of SNR G290.1--0.8 at 1384 MHz as obtained
  with the ATCA. The flux density scale, in units of \mjb, is displayed at 
  the top of the image. The HPBW, $25\farcs0 \times 22\farcs5$, P.A.=$-47\deg$, 
  is indicated in the bottom right corner. The rms noise level is $\sim1$ \mjb. 
  Letters A to H indicate the compact sources discussed in the text. The two 
  grey lines labeled as NE-SW and N-S (both interrupted at the SNR location) 
  show the directions and extent of the two cuts in Fig.~2.}
\end{figure}

\subsection{{\HI} line data}

To construct the \HI \ cube, the continuum component was subtracted in the 
Fourier domain by fitting a linear baseline to 700 line-free channels. The 
data were Fourier-transformed using `natural' weighting, cleaned and restored 
with a synthesized beam of $58\pp \times 45\pp$, P.A.$=-86\deg$. A sensitivity 
of $\sim 5$ \mjb \ was achieved in line free channels. To recover structures 
at shorter spatial frequencies, the ATCA \HI \ data were combined in the 
{\it u-v} plane with data from the Southern Galactic Plane Survey 
\citep[SGPS;][]{sgps}. The latter combinies Parkes and ATCA \HI\ observations. 
No tapering was applied to the lower resolution cube. The line-free channels 
from the combined \HI\ data set have an rms of $\sim$2 K.

For the absorption study, a second \HI \ cube was constructed using the 
method in \citet{rgj+04}, which is described in detail in \citet{dm-cgg03}. 
In this second cube, the continuum was not subtracted and all baselines 
shorter than 1.1 k$\lambda$ were removed. The result is that all \HI \ 
structures larger than $\sim 3.5^\prime$ are filtered out. Since the 
continuum emission from G290.1--0.8 is dominated by smooth, extended 
structures, much of the continuum flux is also filtered out. However, 
the \HI \ absorption features can still be easily seen towards the brightest 
structures. To deconvolve the data, the compact features were cleaned out
and a cube with the residuals was obtained. This cube was further cleaned 
in a second stage. All clean components were restored with a $25\pp \times
23\pp$ beam, with P.A.=$-79\deg$. A filtered continuum image was constructed 
by averaging several line-free channels in the cube, and this was used to 
weight the final data cube. Absorption spectra can be computed by dividing 
the spectrum by the square of the integrated continuum flux from the same 
region.
 
\section{Results}

\subsection{Radio continuum results}

\begin{figure*} 
\includegraphics[width=480pt]{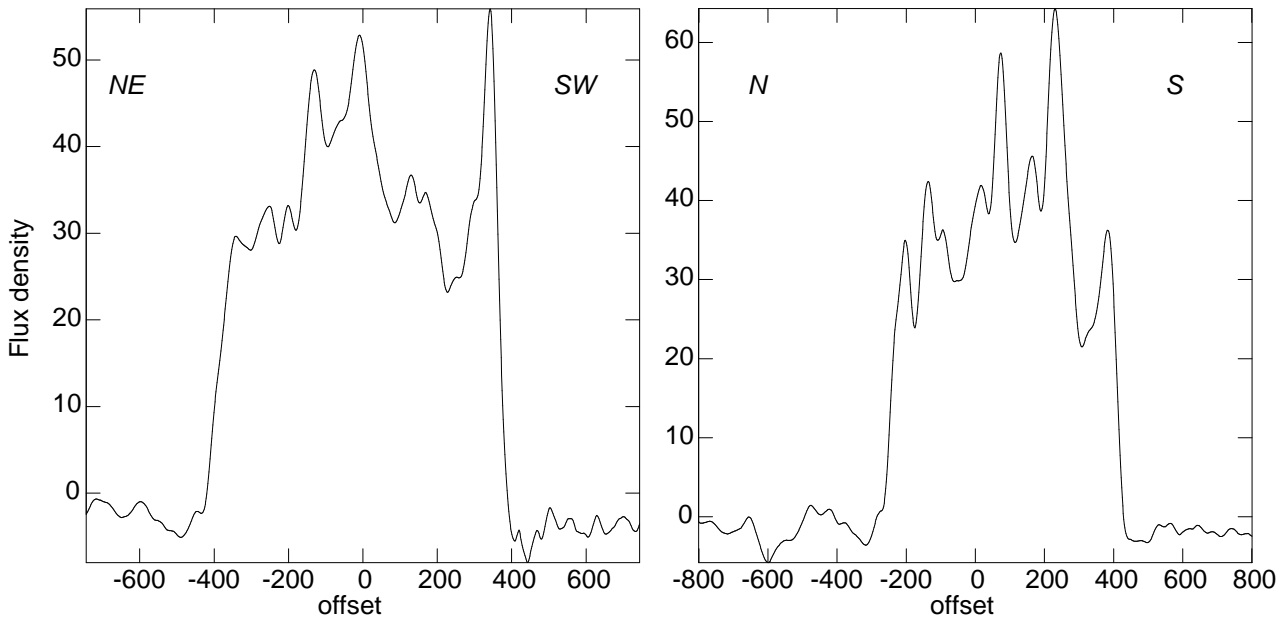}
\caption{Two cuts across SNR G290.1--0.8 in the directions shown in Fig.~1
  ({\it NE-SW} in the {\it left} frame and {\it N-S} in the {\it right}). 
  The flux density is given in \mjb, while the offsets are in arcsec and 
  are referred to the central position in each cut.} 
\end{figure*}

The 20-cm radio continuum image of G290.1--0.8 and the surrounding complex 
region is shown in Figure~1. The SNR shows considerable internal structure. 
The high 
resolution and sensitivity achieved with the present observations reveal a 
network of filaments across the face of the remnant. The sharp southern and 
northern edges of the SNR are remarkably straight, suggesting that the shock 
front may have encountered a plane parallel density gradient. There is some 
evidence for a `blow-out' of material towards the NW and SE. The NW `blow-out' 
ends in a fringe with faint threads oriented in the direction of the major 
axis of the SNR, while the other bulge ends in a thin, very faint tail 
$\sim 4^\prime$ long extending to the south. This weak tail is also seen in 
the 843 MHz image \citep{wg96}.  The flux density of G290.1--0.8 is estimated
to be 25 Jy, which is about 40\% lower than the predicted value based on 
measurements in other frequencies. 

Although this SNR has been classified as shell type, there is no evidence
of shell structure in the 20-cm radio continuum image (Fig.~1) except for 
a bright arc at the western rim. To show this, we plot in Figure~2 the 
intensity along two cuts across the remnant. The directions of the cuts 
(NE-SW and N-S) are shown in Fig.~1. Both cuts show the profiles of multiple 
narrow filaments, perhaps evidence of internal shocks.

The seven strongest compact sources in the field (all with flux densities 
$\geq 30$ \mjb) are labelled A--G in Fig.~1. Source H is a bright patch of
emission projected onto the southern rim of the SNR. Its possible association 
with the remnant is discussed below. Sources B and D are both associated with 
extended continuum emission with total flux densities of 1050 and 100 mJy,
respectively. The parameters for these sources are listed in Table~1. The 
first column corresponds to the label as given in Fig.~1. Some of these 
sources have been identified in SIMBAD, and we have included these 
identifications in the second column of Table~1. To compute the coordinates 
(third and fourth columns) and flux density at 20-cm (fifth column), the
sources have been fitted with two-dimensional Gaussians.
For double sources (see discussion below), the coordinates of the brightest 
component derived using the high resolution image are listed. The sixth column 
gives the computed spectral index. The last two entries (systemic velocity and 
kinematical distance) are discussed in Section~3.2.

To compute the spectral indices, we used an 843 MHz radio continuum image 
(HPBW of $52\pp \times 45\pp$) obtained from the Sydney University Molonglo 
Sky Survey \citep[SUMSS;][]{sumss}. The ATCA 20-cm radio continuum image was 
convolved to the same resolution as the Molonglo image, and fluxes at both 
frequencies were compared in such a way that the slope between them gives a 
measure of the spectral index \citep{turtle}. The method requires that the 
two images to be compared are filtered for the same {\it u-v} coverage 
\citep{gbm+99}. Although this condition was not rigorously fulfilled here, 
the method may be used to give a good approximation for structures smaller 
than a few arcmin.

\begin{table*} 
\caption{Parameters of the compact sources near SNR G290.1--0.8}
\begin{tabular}{llccrccc}
\hline
\hline
Source & Identification & Right & Declination & Flux \ \ &Spectral &
Systemic & Kinematical \\
      &  & Ascension  & & &index  & velocity &
distance \\
& & (J2000)   & (J2000)   \\ 
& & h m s & $\circ \ \ \prime \ \ \prime\prime$  &(mJy) && (\k) & (kpc) \\
\hline
A     &  & 11 04 55        & --60 45 58  &   77    & $-1.1\pm 0.1\fant$ &
$> +77$ & $> 14$  \\
B     & G289.8--0.8, & 11 00 59        & --60 50 23  &  507    & 
$0.63\pm 0.08$ & (+20 , +28) & (7.7 , 8.4)    \\
&IRAS 10588--6030,\\
&[CH87] 289.878--0.792,\\ 
&MSH 11--6{\it 1}B\\
C     &  & 11 04 39        & --61 09 27  &  337    & $-0.85\pm 0.10$ &
$> +115$ & $> 20$    \\
D     & G289.95--0.89, & 11 01 11  & --60 56 49  & 82 & $\fant 0.0\pm 0.1\fant$
& (+31.5 , +70) & (8.8 , 13)     \\
& IRAS 10591--6040\\
E     &  & 11 01 34        & --60 53 08  &   35    & $-0.6\pm 0.2\fant$  &
$> +75$  & $> 13$   \\
F     &  & 11 01 58        & --60 56 50  &   57    & $\fant 0.5\pm 0.1\fant$    
& $> +80$ & $> 14$ \\
G     &  & 11 01 41        & --60 46 45  &   27    & $-0.7\pm 0.1\fant$  &
$> +75$  & $> 13$   \\
H     &MSX G290.2181-00.8153, & 11 03 28 & --60 59 33  &   40 & $-0.4\pm 
0.3\fant$ & $> +80$ & $> 14$ \\
&IRAS 11014--6044\\
\hline
\end{tabular}
\end{table*}

The high resolution image reveals that several of these sources are extended
or double. Source A shows a secondary peak more than 3 beams to the north of
the brightest component, with a flux lower by a factor of \ga 2. Source B is
resolved, and only $\sim 18\pp$ away from its centre there appears a secondary
source one order of magnitude fainter. Source C is resolved into two extended 
components separated by $\sim 45\pp$, one of them being 4 times brighter that
the other. Source H appears as a faint filament of size $\sim 16\pp \times 4 
\pp$ projected against the SNR. The two pulsars in the area were both detected. 
PSR J1104--6103's position is consistent with timing observations 
\citep{kmj+96}. There is a 4$\sigma$ detection of the second pulsar, 
PSR J1105--6107. 

\begin{figure} 
\includegraphics[width=252 pt]{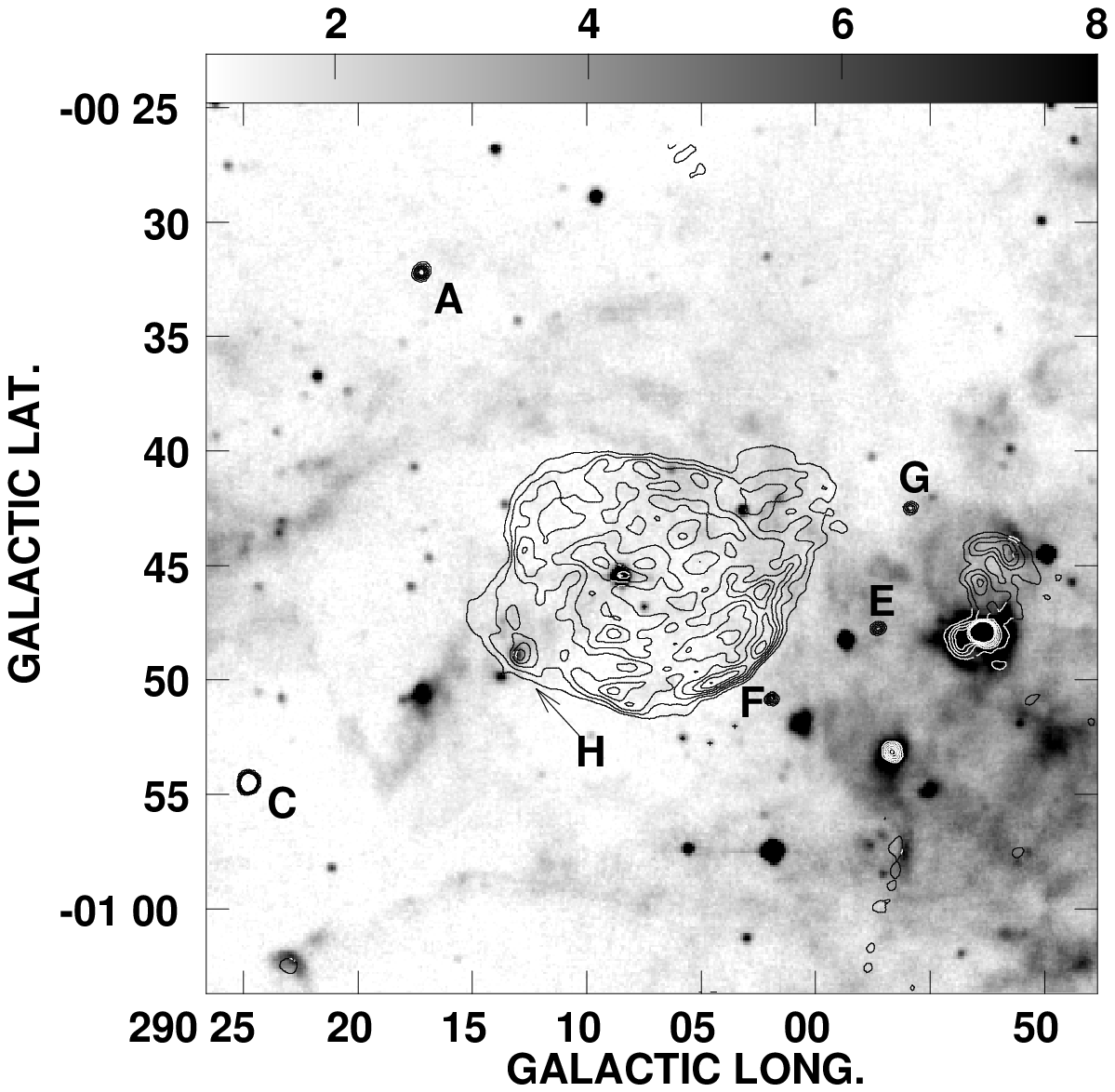}
\caption{MSX infrared emission towards SNR G290.1--0.8 at 8.3 $\mu$m in 
  grayscale with 20-cm radio continuum contours overlaid. The grayscale, 
  in units of $10^{-6}$ W/m$^2$ sr, is indicated at the top of the image. 
  The contour levels range from 0 to 70 \mjb, in steps of 10 \mjb.
  For clarity, white contours are used over dark greyscale background 
  (e.g., sources B, D, H, and IRAS 11010--6038 at the centre of the SNR). 
  Source H (see Fig.~1) is indicated with an arrow. Labels have been 
  included for sources A, C, E, F, and G to avoid confusion between black 
  radio contours and MSX emission.}
\end{figure}

We have also examined the  8.3 $\mu$m infrared image from the Midcourse Space 
Experiment (MSX)\footnote{http://irsa.ipac.caltech.edu/applications/MSX/MSX/},
shown overlaid onto our radio continuum image of G290.1--0.8 in Figure~3. A 
few infrared point sources can
be seen within the outermost contour of the SNR. The strongest of these
(G290.1423-00.7577 in the MSX6C catalogue, or IRAS 11010-6038) is the 
infrared counterpart of the variable star dubbed as HD 95950 or V0 528 Car,\
with a spectral type M2Iab. Source H lies slightly off-center (by $\sim 5\pp$)
from an IR source catalogued as MSX G290.2181--00.8153 and IRAS 11014--6044. 
The faint protrusion to the south of the SNR near H (see also Fig.~1), is 
found to be another unresolved source coincident with the infrared sources 
G290.2292-00.8311 (MSX6C catalogue) and IRAS 11014-6044.

\subsection{{\HI}\ results}

The tangent point at $l=290^\circ$ is located at a distance of 2.9 kpc.
According to the Galactic rotation model of \citet{fbs89}, there should be 
no \HI \ gas with velocities more negative than $-12$ \k . However, we
observe \HI \ emission to --30 \k. \citet{jkww96} also detect departures
in velocity from the Galactic rotation model of up to 20 \kms in this
direction. The Carina spiral arm lies virtually along the line of sight 
at this longitude \citep{gg76}, substantially affecting the gas velocities 
in this part of the Galaxy.

\subsubsection{Compact sources}

\begin{figure*}
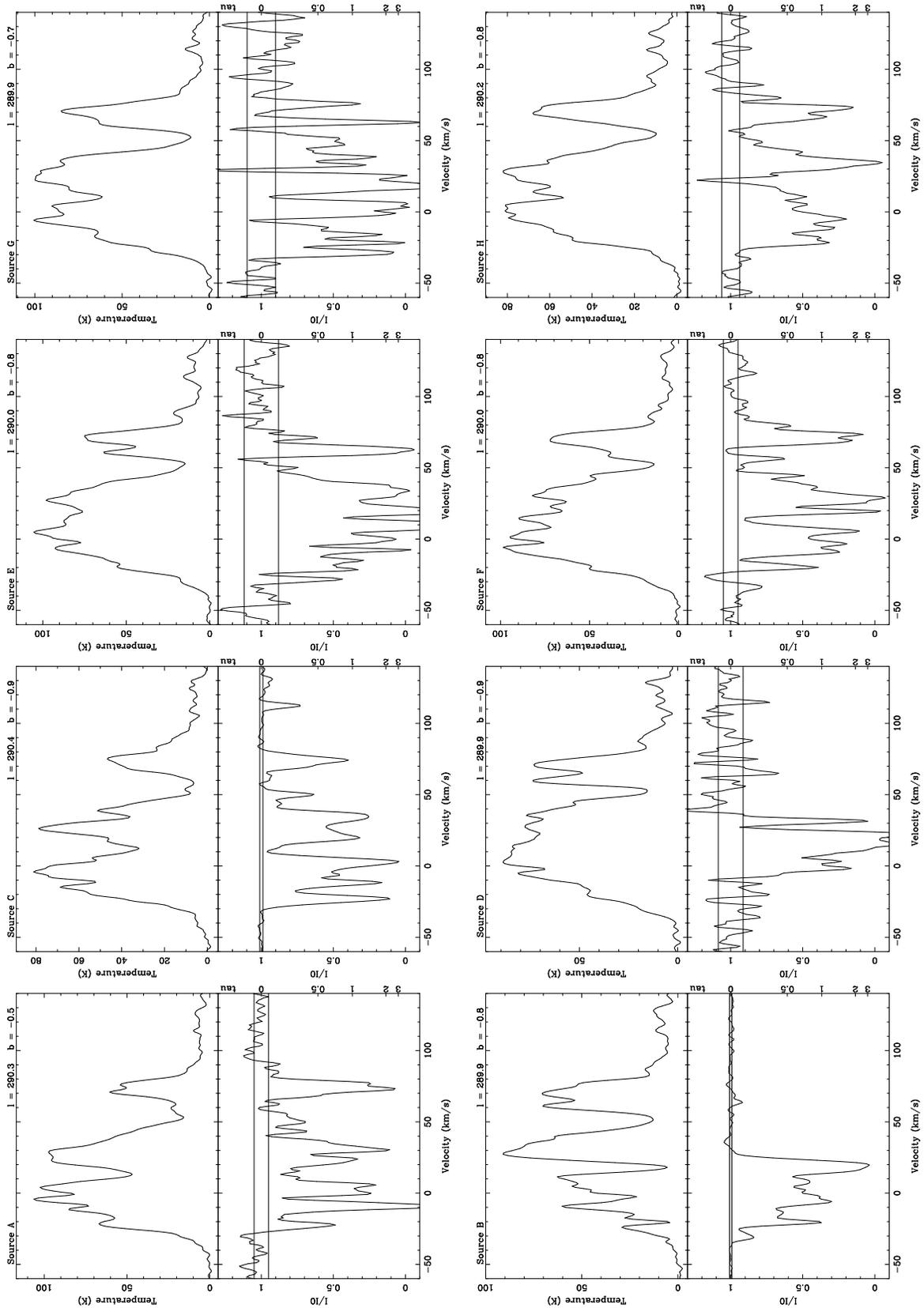
 
\vbox to 220mm{\vfil
\begin{tabular}{cc}
\includegraphics[width=5.4 cm,angle=90]{Fig4-G.ps}&
\includegraphics[width=5.4 cm,angle=90]{Fig4-H.ps}\\
\includegraphics[width=5.4 cm,angle=90]{Fig4-E.ps}&
\includegraphics[width=5.4 cm,angle=90]{Fig4-F.ps}\\
\includegraphics[width=5.4 cm,angle=90]{Fig4-C.ps}&
\includegraphics[width=5.4 cm,angle=90]{Fig4-D.ps}\\
\includegraphics[width=5.4 cm,angle=90]{Fig4-A.ps}&
\includegraphics[width=5.4 cm,angle=90]{Fig4-B.ps}\\
\end{tabular}
\caption{\HI\ brightness temperature profiles ({\it top}) and absorption 
  spectra ({\it bottom}) towards the compact sources in the field. The \HI\ 
  profiles, in K, are from the cube that combines Parkes and ATCA data. 
  The absorption spectra and the corresponding emission profile, were computed 
  as described in the text. In the absorption spectra, the vertical axis shows
  the ratio between the intensities at each velocity channel and the continuum 
  (I/I$_0$, or e$^{-\tau}$) on the {\it left} and as opacity, $\tau$ on the
  {\it right}. All profiles were Hanning-smoothed. The name of each source and 
  their Galactic coordinates are indicated at the top of each \HI \ profile. 
  The two horizontal lines close to I/I$_0 =1$ show the $1\sigma$ noise level 
  for each absorption profile. All velocities are with respect to the LSR.}
\vfil}
\end{figure*}

\begin{figure} 
\includegraphics[width=252 pt]{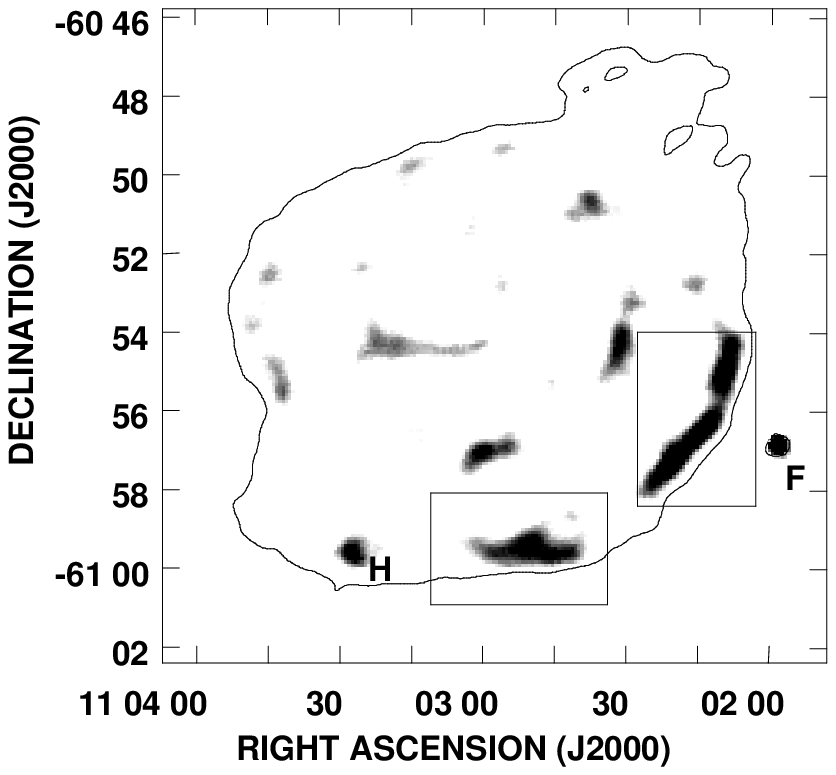}
\caption{20-cm radio continuum image of SNR G290.1--0.8 constructed by 
  averaging several line free channels of the filtered \HI\ cube. Greyscale 
  shows the filtered intensity remaining after pixels below $5\sigma$ 
  were blanked. The noise level is 3 \mjb. Boxes indicate the regions 
  where \HI\ absorption spectra are computed (south and west). 
  The location of sources F and H are also indicated. A 20-cm radio 
  contour at 1 \mjb \ is included.}
\end{figure}

Spectra towards all sources (A -- H) were obtained in both cubes in order 
to compare \HI \ emission and absorption. All profiles are displayed in 
Figure~4. The name of each source and their Galactic coordinates are 
indicated at the top of the \HI \ emission profile, while the corresponding 
absorption profile is shown below. A brief description follows of the spectra 
for each of the sources. 

Source A shows strong absorption up to +77 \k, coincident with the
most positive velocity emission feature, which implies a lower distance 
limit of 14 kpc. The steep spectral index indicates an extragalactic
origin for this source.

Source B, named G289.9--0.8 by \citet{sg70b} and also known as 
MSH 11--6{\it 1}B, shows strong absorption through all negative velocities 
and at +20 \k. We assign source B a lower distance limit of 7.7 kpc and an 
upper limit of 8.4 kpc, the latter corresponding to +28 \k, which is the first 
emission feature with no associated absorption. \citet{bnm96} classify this 
source as an ultracompact \HII \ region based on the detection of the CS 1-0 
line at about +20 \k . The extended source possibly associated with source B 
is very faint at 20-cm, which makes an absorption study difficult. From the 
\HI \ data it is not possible to establish a physical link between the compact 
source B and the extended emission. The spectral index of the diffuse emission 
to the north of source B is $0.0\pm 0.1$, typical of \HII \ regions, while the 
spectral index of source B is $+0.63\pm 0.08$, which is expected for stars with 
an ionized, uniform, spherically symmetric mass loss flow \citep{wb74}.

Source C, shown as G290.4--0.9 on early radio images, shows strong absorption 
features to +115 \k. This is the only source to clearly show absorption
at such a high velocity. However, the brightness temperatures for the other
sources in Table~1 are lower and any high velocity absorption may be lost in 
the noise. The high resolution image which shows that the source is a compact 
double and the steep spectral index strongly indicate that the source is 
extragalactic.

Sources D, E and F are all located within 6 arcmin of each other. The 
emission spectra in the direction of these sources are similar and yet D (the 
strongest of the 3 sources) shows a rather different absorption spectrum. 
Sources E and F clearly show absorption to +33 \kms and again against the
emission feature near +70 \k , which implies lower distance limits of 
approximately 13 and 14 kpc respectively. In contrast, source D shows no 
evidence for absorption against the +70 \kms feature and is therefore likely to 
be a Galactic source. The lower distance limit to D is 8.8 kpc, based on the 
absorption feature at +31.5 \k. Furthermore, it appears to be resolved in the 
visibilities measured for the longest baselines, recording only 25\% of the
total intensity. Source D also has strong infrared emission and is coincident 
with a water-vapor maser \citep{scalise} located 8.9 kpc away, derived from 
a systemic velocity of $\sim +33$ \kms from CS and C$^{34}$S observations 
\citep{zmt95}.

Source G is the weakest point source listed. Its absorption spectrum is
rather noisy and difficult to interpret. There appears to be absorption out 
to +75 \k, implying a lower distance limit of $\sim 13$ kpc. Finally, source 
H clearly shows absorption to +80 \k, which sets a lower limit of 14 kpc for 
the distance. The range of possible velocities for each source according to 
the discussion above, and the derived distances, are summarized in the last 
two columns of Table~1.

\subsubsection{Supernova remnant}

To determine the kinematical distance to G290.1--0.8, we analyzed absorption
profiles towards different regions of the SNR, using the procedure described
in Section~2.2. Pixels below $5\sigma$, where $\sigma=3$ \mjb, have been 
blanked. For each absorption spectrum, we produced 5$\sigma$ envelopes to
determine which features were significant. This method also filters out smooth 
scale continuum emission. From Figure~5, where the resulting filtered
continuum image is shown, it is seen that there are not many areas of the 
SNR where there is an opportunity to prepare absorption spectra using this
method.

\begin{figure} 
\includegraphics[width=252pt]{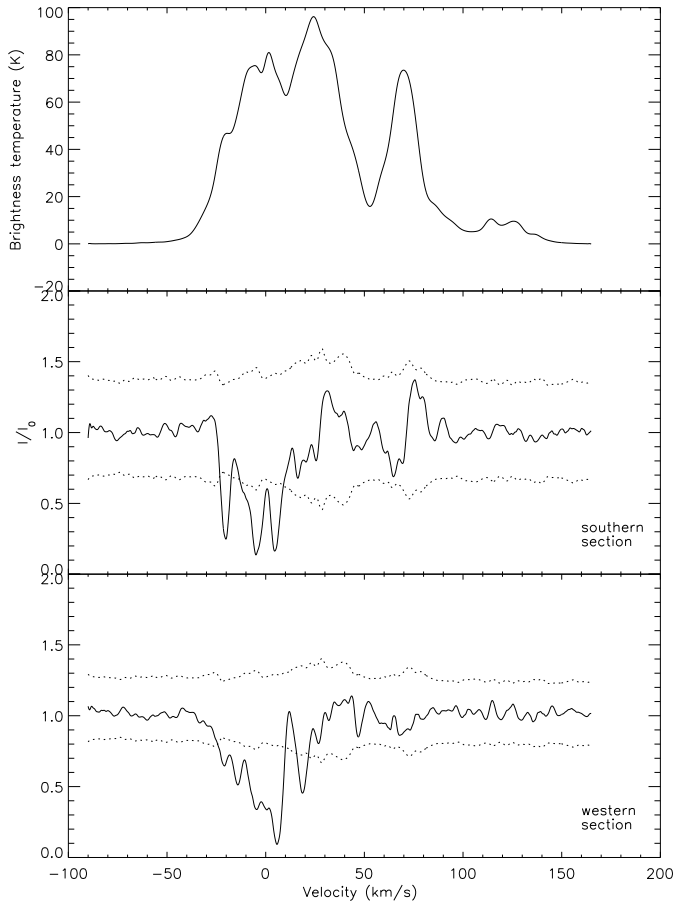}
\caption{Upper frame: average \HI\ emission profile towards SNR G290.1--0.8.
  Lower frames: \HI\ absorption spectra towards two regions, to the south and 
  to the west, as shown by the boxes in Fig.~5. 
  The envelopes indicate 5$\sigma$ uncertainties.} 
\end{figure}

In Figure~6, the two most convincing absorption profiles are displayed, 
towards the south and the west of the SNR (shown with boxes in
Fig.~5). An average \HI\ emission spectrum towards the whole area 
subtended by the SNR is included. The two absorption profiles look quite
different: the western spectrum shows no absorption beyond about +20 \k, 
while the southern spectrum shows no absorption beyond about +7 \k.

The interpretation of these \HI\ spectra, particularly with respect to
estimating the distance to the SNR, is complicated and described in the
next section.

\section{Discussion}

\begin{figure*} 
\includegraphics[width=470pt]{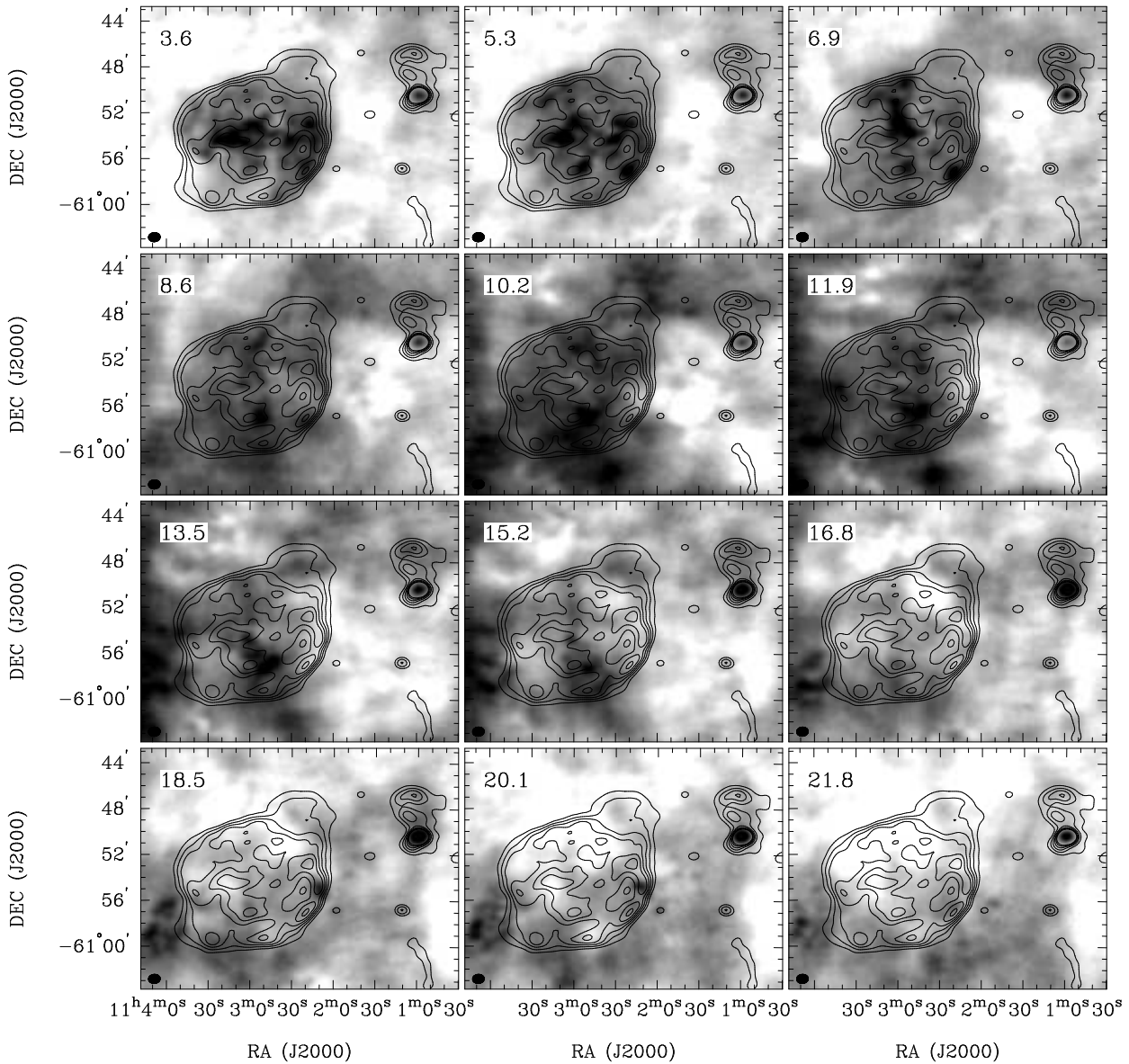}
\caption{\HI\ channel maps (greyscale) of the ATCA + SGPS combined data cube 
  covering SNR G290.1--0.8 and source B over the velocity range from +3.6 to 
  +21.8 \k, overlaid on the smoothed 20-cm radio continuum map (contours). 
  The contour levels are 30, 70, 110, 150, 190, 230 \mjb. The greyscale ranges 
  from 40 K (dark) to 100 K (bright). The beam ($59\pp \times 45\pp$, P.A.= 
  $-86^\circ$) is displayed in the bottom left corner of each panel. For 
  display purposes we chose a channel width of 1.65 \k, twice the original 
  width. The \HI \ velocities are indicated at the top left corner of each 
  panel.}
\end{figure*}

A visual inspection of the \HI\ data cube shows that SNR G290.1--0.8 is 
seen unambiguously in absorption for velocities from --30 to approximately 
+7 \k. Beyond this velocity the picture is not clear. 

Figure~7 shows images at 12 velocity channels from the \HI \ cube to demonstrate
the complexity of the structures and illustrate the difficulty in determining 
the systemic velocity for this SNR. \HI \ absorption is clearly seen against 
most of the remnant out to velocities of about +7 \k. The first channels 
displayed in Fig.~7 clearly show an ``\HI \ shadow'' within the outer contours 
of the radio continuum emission. We note that the \HI \ emission and absorption 
across the face of the remnant vary substantially, with absorption structure 
that often does not correlate well with the continuum intensity peaks. Beyond 
about +7 \kms there is no obvious ``\HI \ shadow'' at the position of the 
remnant, partly due to diminished \HI \ emission at velocities up to about 
+14 \k. A slight drop in \HI \ emission at the western edge of the remnant,
at velocities around 18--20 \k, is difficult to interpret, but most likely not 
due to absorption. In contrast, \HI \ absorption against the very bright radio 
continuum emission of source B is clearly seen out to about +24 \k.

The \HI\ spectrum against the western rim (Figure~6) shows a local minimum 
at +20 \k. This is one of several features which are marginally above the 
$5\sigma$ detection limit. The feature does not appear in the \HI\ spectrum 
against the southern rim. It is not possible to determine conclusively from 
the present data if this feature is associated with absorption by the SNR. 
If this velocity does represent the systemic velocity of the remnant, it 
would correspond to a distance of $7.7 \pm 0.7$ kpc.

Another anomaly complicating the interpretation of G290.1--0.8 concerns the
bright compact source H which is located within the SNR rim. A comparison
with the clearly unrelated source F is helpful. Both sources are shown in 
Figure~5. Inspection of the two absorption spectra (Figure~4) shows obvious
similarities, with strong positive velocity features 
at +33 \kms and +70 \k, indicating distances beyond the solar circle of
more than 14 kpc.  Source H also appears to have an infrared
counterpart, catalogued in both the IRAS (11014--6044) and MSX
(G290.2181--00.8153) surveys. Furthermore, the weak finger of continuum
emission extending from source H, outside the SNR rim, also has an MSX
source at the same position. Hence, it seems most likely that
source H is unrelated to the SNR and the shock heated dust model to
explain the IRAS spectrum \citep{junkes} has an origin not related to
the formation of the SNR.

A common problem when studying \HI \ absorption against a continuum
source is confusion with self-absorption against the \HI \ background. 
This happens when the optical depth along the line of sight is high and 
the phenomenon is typically seen as narrow features indicating the presence 
of small, cold clouds. For a small scale object, an interferometer observes 
a brightness temperature $T_v=T_s - p T_g$ \citep{sgkb95}, where $T_s$ is 
the \HI\ spin temperature, $T_g$ is the background \HI \ emission and
$p$ is the fraction of absorbed \HI. To test the nature of the
observed absorption, we followed the method employed by \citet{sgkb95}
to determine the distance to Tycho's SNR: we plotted the minimum \HI \ 
value per channel in the unfiltered interferometric cube towards the 
whole area subtended by G290.1--0.8, and towards regions away from the 
SNR in the four directions, north, south, east and west, all covering the 
same area. If features are similar on- and off-source, then self-absorption 
may be the explanation. The spectra are shown in Figure~8. 
Clear absorption against the continuum emission from G290.1--0.8 is 
not found much beyond +7 \k. The possibility of the +20 \k\ feature 
being associated with the SNR is discussed previously. 
From Figure~8, we can deduce that absorption is definitely associated out 
to +7 \k, by observing the large differences between the SNR spectrum 
and the other spectra. For more positive velocities we again see inconsistent 
results. Adopting a value of +7 \kms for the systemic velocity, the
corresponding distance is about $7 \pm 1$ kpc, using the Fich et al. (1989) 
model for Galactic rotation. Note that the distance derived previously for
+20 \k\  is not greatly different from our proposed estimate and likely to 
be consistent within the uncertainties of the model and expected 
peculiar velocities.

While the centrally peaked X-ray source is located near the central bright
filament of the SNR, the distance estimate from the X-ray data of 8 -- 11
kpc is not sufficiently well-constrained to help us determine the velocity
range of features that are associated with absorption by the SNR. The
column density measured by integrating the \HI \ from the most negative
velocity to the highest related positive velocity (taken as +10 \k) is
$N_H = 4.5 \times 10^{22}$ \cm2, which is not inconsistent with the value 
estimated by \citet{sshp02} of $1.3 \times 10^{22}$ \cm2, given the 
uncertainties. The values would be even closer if the velocity of +20 \kms 
is adopted as the systemic velocity. 

On balance, it seems likely that the 
systemic velocity is close to +10 \k, which is in good agreement with 
the H$\alpha$ result from \citet{ralm96} of +12 \k, corresponding to a 
kinematic distance of $\sim 7$ kpc. This places the SNR in the Carina arm 
with a linear diameter of about 25 pc at a scale-height of nearly 100 pc, 
which are consistent with parameters for the known SNR population. The 
results from this work do present a consistent picture, notwithstanding 
the difficulty in interpreting the highly complex structure seen in \HI \
data, which can be due not only to absorption by continuum sources, but 
real density variations, turbulence and velocity structure and 
self-absorption by cold clouds in the ISM.

\begin{figure} 
\includegraphics[width=252pt]{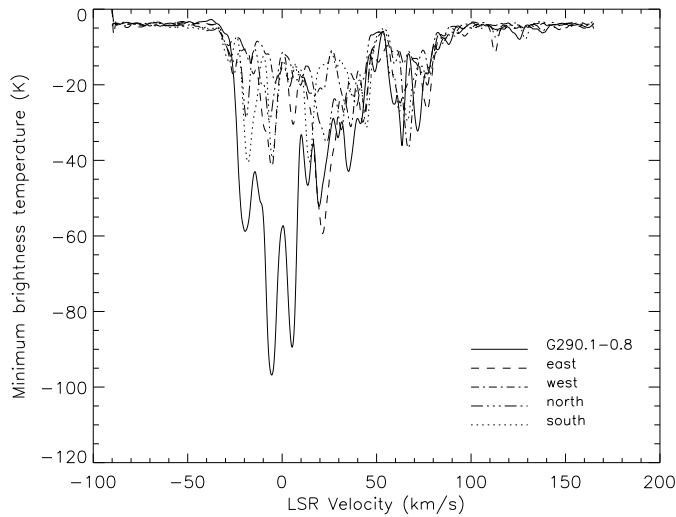}
\caption{Spectra of the minima in the \HI\ data as measured on the 
unfiltered ATCA data cube towards five regions of similar area:
one over SNR G290.1--0.8, and four over the surroundings.}
\end{figure}

The surface brightness of G290.1--0.8 is $\sim 3.5\times 10^{-20}$
W$^2$ m$^{-2}$ Hz$^{-1}$ sr$^{-1}$. Statistically, high surface 
brightness SNRs (above $1\times 10^{-20}$ W$^2$ m$^{-2}$ Hz$^{-1}$ 
sr$^{-1}$) are strongly concentrated around $l=0^\circ$, but there
is another peak at $l=290^\circ$ \citep{green04}. If this secondary
peak is due to the fact that in this direction the Carina arm is
crossed tangentially, and there is a concentration of high surface
brightness SNRs in this arm, then it is very probably that G290.1--0.8
also belongs to the Carina arm. This is supporting evidence that 
G290.1--0.8 is at a distance of 7 kpc (systemic velocity $\sim +10$ \k). 

\section{Conclusions}

We have imaged the SNR G290.1--0.8 in both the radio continuum and the 
21-cm \HI\ line using the ATCA. To recover the shortest spatial frequencies, 
our \HI\ observations were combined with SGPS data (McClure-Griffiths et
al. 2005). Absorption was analyzed by filtering the \HI\ data, keeping only 
the highest spatial frequencies (above 1.1 k$\lambda$). 

In the 20-cm radio continuum image, G290.1--0.8 shows filamentary emission 
with little evidence of shell structure, although this SNR is classified
as a shell-type.  The northern and southern edges of this SNR are
remarkably straight, suggesting that the shock front may have
encountered a plane parallel density gradient. However, no emission
features with similar morphology were found in the \HI \ cube at any 
velocity likely to be associated with G290.1--0.8.

The \HI\ in this direction is very complex since the line of sight 
cuts tangentially through the Carina spiral arm. Departures in velocity 
of $\sim 20$ \kms with respect to circular rotation model are common. 
To resolve the problem of the kinematic distance to G290.1--0.8, we 
made various tests and conclude that the upper limit of the systemic 
velocity is about +7 \k, also in good agreement with the result obtained 
from H$\alpha$ data \citep{ralm96}. This velocity is translated into 
a distance of $7 \pm 1$ kpc, in good agreement with the lower limit 
inferred from X-rays ASCA data \citep{sshp02}.

Additionally, we studied eight radio sources in the field, including
a compact knot in the southern rim of G290.1--0.8 which is probably
an unrelated source. A high resolution image constructed using only 
the baselines involving the 6th ATCA antenna (baselines from 2.6 to 
5 km) revealed that some of these sources are close doubles. Spectral 
indices were computed using the flux-flux method combining our 
observations with data from the SUMSS at 843 MHz. Distance limits 
were derived for all these sources based on their absorption spectra.

\medskip

\section*{Acknowledgments}

The authors wish to thank Naomi McClure-Griffiths for her advices on the
\HI \ absorption analysis.
This project was partially financed by grants ANPCyT-14018 and UBACYT A055
(Argentina). During part of this work, E. M. R. was a visiting scholar at 
the University of Sydney. The ATCA is part of the Australia Telescope, which 
is funded by the Commonwealth of Australia for operation as a National Facility 
managed by CSIRO. This research has made use of the NASA/ IPAC Infrared Science 
Archive, which is operated by the Jet Propulsion Laboratory, California 
Institute of Technology, under contract with the National Aeronautics and Space 
Administration.

\label{lastpage}

\end{document}